\documentclass[prl, showpacs, superscriptaddress, twocolumn, floatfix, amsmath]{revtex4}
\usepackage{color, graphicx, amssymb}  

\date{09/07/07}
  
\begin{document}  
\title{Multiple fluxoid transitions in mesoscopic superconducting rings}  
\author{Hendrik Bluhm}  
\email{hendrikb@stanford.edu}  
\affiliation{Departments of Physics and Applied Physics, Stanford  
University, Stanford, CA 94305}  
\author{Nicholas C. Koshnick}  
\affiliation{Departments of Physics and Applied Physics, Stanford  
University, Stanford, CA 94305}  
\author{Martin E. Huber}  
\affiliation{Department of Physics, University of Colorado at Denver,   
Denver, CO 80217}  
\author{Kathryn A. Moler}  
\affiliation{Departments of Physics and Applied Physics, Stanford  
University, Stanford, CA 94305}  

\begin{abstract} The authors report magnetic measurements of fluxoid
transitions in mesoscopic, superconducting aluminum rings.  The
transitions are induced by applying a flux to the ring so that the
induced supercurrent approaches the critical current. In a temperature range
near $T_c$, only a single fluxoid enters or leaves at a time,
leading to a final state above the ground state. Upon lowering the
temperature, several fluxoids enter or leave at once, and the final state
approaches the ground state, which can be reached below approximately
0.5 $T_c$.  A model based on the widely used time dependent
Ginzburg-Landau theory for gapless superconductors can only explain the
data if unphysical parameters are used. Heating and quasiparticle diffusion 
may be important for a quantitative understanding of this
experiment, which could provide a model system for studying the nonlinear 
dynamics of superconductors far from equilibrium.
\end{abstract}

\pacs{74.40.+k, 74.78.Na}
% 74.40.+K Fluctuations (noise, chaos, nonequilibrium superconductivity,
%     localization, etc.)
% 74.78.Na Mesoscopic and nanoscale systems

\maketitle

The dynamics of
phase slips (PS) in quasi-1D superconducting wires has long been  
studied intensively. Most observed phenomena have been explained with
phenomenological models \cite{SkocpolWJ:Phacan, KadinAM:Chaiwa} 
and the time-dependent Ginzburg-Landau theory (TDGL) 
for gapped superconductors \cite{KramerL:Thedcs}, often with good 
quantitative agreement \cite{TidecksR:Curnpi}.
However, in those wires with diameter $d$ of typically a few $\mu$m,
the quasi-1D limit $\xi(T) \gtrsim d$ and  $\lambda(T) \gg d$ was
only accessible very close to the critical temperature $T_c$, 
where the coherence length $\xi$ and penetration depth $\lambda$ diverge.
Due to various approximations, the microscopic validity of most of the 
theoretical models employed is also limited to the vicinity of $T_c$. 
Modern e-beam and nanotemplating techniques yield 
superconducting wires 
which remain quasi-1D down to $T = 0$, and can be thin enough ($d \sim$ 10 nm)
to study phase transitions due to quantum PS's.
Such quantum phenomena, which have recently attracted tremendous 
interest \cite{LauCN:Quapss}, are possible when some action 
$S \sim \Delta E \Delta t$ is comparable to $\hbar$. 
Thus, the dynamical laws determining the time scale $\Delta t$ are
very important for understanding quantum effects.  Furthermore, a
better understanding of superconducting dynamics is interesting both
for its own sake and for device applications, such as microbridge
SQUIDs \cite{HasselbachK:Micqid} or nanowire single photon detectors
\cite{AnnunziataAJ:Supnns}.

We have probed the dynamical evolution of the order parameter in
mesoscopic Al rings at temperatures $T = 0.15 \dots 0.9~T_c$.
Upon increasing the magnetic flux $\Phi_a$ threading a ring, the circulating 
supercurrent $I$ increases quasistatically until it approaches the 
critical current, and the ring switches rapidly to a more stable state.
We find that near $T_c$, the phase winding number $n$ of the order parameter
$\psi(x) = |\psi| e^{inx}$ changes by $\Delta n$ = 1 at each transition, 
which can leave the ring in a metastable state well above the ground 
state. At lower $T$, $\Delta n$ increases, and the final state approaches 
the ground state, while static states persist over a larger range of 
applied flux.
Below $T \approx 0.5~T_c$ , the ring decays into the ground state over an
increasing range of flux bias points. Such multiple fluxoid transitions 
were predicted based on the gapless TDGL equations \cite{VodolazovDY:Dyntbm}
and observed at a single $T$ \cite{VodolazovDY:Mulfja}.
Refs. \cite{BergerJ:Flutsr} and \cite{TarlieMB:Metsso} contain further 
related theoretical work on state selection and the effect of fluctuations 
on the dynamics of mesoscopic superconducting rings, albeit in a different 
parameter range.
The TDGL model in \cite{VodolazovDY:Mulfja,VodolazovDY:Dyntbm}
can only reproduce our and previous ring data by choosing 
a numerical constant which is inconsistent with 
microscopic theory and transport measurements in the resistive state. 
The gapped TDGL theory may 
overcome this inconsistency. However in Al, both versions of TDGL are 
justified only extremely close to $T_c$, 
and we estimate that the neglected heating and quasiparticle (QP) diffusion
may be important in our experiment.
Lacking a tractable theory for superconducting dynamics 
far from $T_c$, it is common to use TDGL, sometimes  
well below $T_c$ and in the gapless version. 
Thus, it is of considerable interest to understand to what extent
those theories are adequate outside their 
theoretical range of validity, and to develop better theories. 
Experiments such as ours can be used for validation.

Unlike transport measurements on superconducting wires, where
voltage reflects the mean phase winding rate, our experiment
probes when phase unwinding ceases, once started.
Thus, we explore fast dynamics without a high 
measurement bandwidth.
Furthermore, DC transport measurements are prone to 
thermal runaway effects at low $T$, where the critical current is large 
and cooling is inefficient.
In our case, heating is also significant, but 
localized in space and time due to the finite number of PS's.
The absence of leads eliminates the need to consider proximity effects 
and boundary conditions at the contacts.

\begin{figure}
\includegraphics{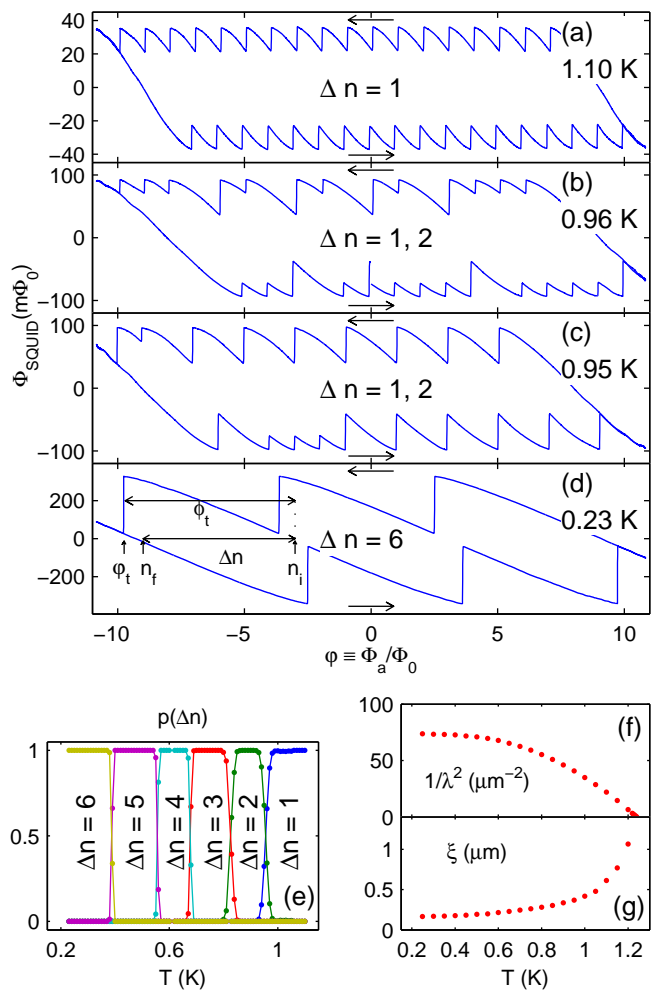}
\caption{\label{fig:sweeps}\label{fig:prob} 
(Color online) (a)-(d)
Typical single-sweep $\Phi_a$--$I$ curves for a $R = 2 \mu$m, 
$w = 180$ nm ring. The ring current $I$ is proportional to the SQUID response
$\Phi_{SQUID}$, with mutual inductance $M_{coup} \approx 1.2~\Phi_0/$mA for 
this ring size.
At 1.1 K (a), $\Delta n = 1$ always. Around 0.96 K (b) and 0.95 K (c),
$\Delta n=2$ transitions become increasingly likely. At 0.28 K (d),
$\Delta n = 6$, and the ring decays close to the ground state at each 
transition. The arrows indicate the field sweep direction and illustrate 
$\Delta n$, $n_i$, $n_f$ and $\phi_t$ for the transition 
at $\varphi_t = -9.7$.  
(e) Probabilities for a change $\Delta n$ in the 
fluxoid number. (f),(g) $\lambda^{-2}$ and $\xi$ as a function of $T$.
}
\end{figure}

The results reported here are from 13 rings with radii 
$R$ = 1 and 2 $\mu$m and line width $w$ = 70, 115, 140 and 180 nm, 
measured in a single cool down. 
We measured $I$ as a function of $\Phi_a$ and $T$ by
positioning a scanning SQUID microscope \cite{BjornssonPG:Scasqi} 
over each ring individually.
The samples were made by e-beam lithography using PMMA resist and liftoff.
The 60 nm thick Al film was deposited using 
e-beam evaporation at a pressure of 10$^{-7}$ mBar and a rate 
of 35 \AA/s.  Using $\xi_0 = 1.6~\mu$m for pure bulk Al
\cite{MeserveyR:Equpce} and $\xi(0) =$ 140 to 180 nm depending on $w$,
obtained as described below [see also Fig. \ref{fig:prob}(g)],
we infer a mean free path of $l_e = 1.4 \xi(0)^2/\xi_0 =$ 17 to 28 nm. 
The measured $T_c$ is 1.25 K. 

Our SQUID sensor is designed as a susceptometer as discussed in 
\cite{BluhmH:Magrms,GardnerBW:Scasqi}. 
We positioned one of its pickup loops over a ring and recorded 
its response 
while sinusoidally varying $\Phi_a$, generated by an on-chip field coil, 
at 0.6 Hz.To isolate the flux generated by $I$, we subtracted a background 
measured by retracting the SQUID from the sample.
For each ring, we 
took 40 such $\Phi_a$--$I$ curves at each $T$, in steps
of 10 mK --- a total of more than 40,000 field sweeps containing 
550,000 transitions.

We also collected data at a smaller $\Phi_a$-amplitude, so that 
fluxoid transitions only occurred near $T_c$, to extract
$\lambda^{-2}(T)$ and $\xi(T)$ [Fig \ref{fig:prob}(f, g)],
as described in Ref \cite{BluhmH:Magrms}.
Using fifth order rather than cubic polynomials to 
fit the  low $T$ $\Phi_a$--$I$ curves accounts for corrections to 
1D GL at low $T$\cite{AnthoreA:Denssc}, and for
imperfections in the rings, whose effect is biggest at large $I$.
We also corrected for partial screening of the applied flux.

Fig. \ref{fig:sweeps} shows a few individual field sweeps. 
Each branch of a $\Phi_a$--$I$ curve represents the response 
$I_n(\varphi \equiv \Phi_a/\Phi_0)$ of a
state with fluxoid number $n$, with a linear response around 
$\varphi = n$, and a curvature caused by pair breaking at larger 
$\varphi-n$. $\Phi_0 = h/2e$ is the superconducting flux quantum.
Since $w \ll R$ and $H_a \ll H_{c2}$, $I_n$ and the transition points only 
depend on $\varphi-n$ to high accuracy. In the GL regime, 
$I_n(\varphi) = -(wd \Phi_0/2 \pi R \mu_0 \lambda^2) (\varphi-n)  
(1-(\xi^2/R^2) (\varphi-n)^2)$. At lower $T$, deviations from the cubic form  
become noticeable.

\begin{figure}
\includegraphics{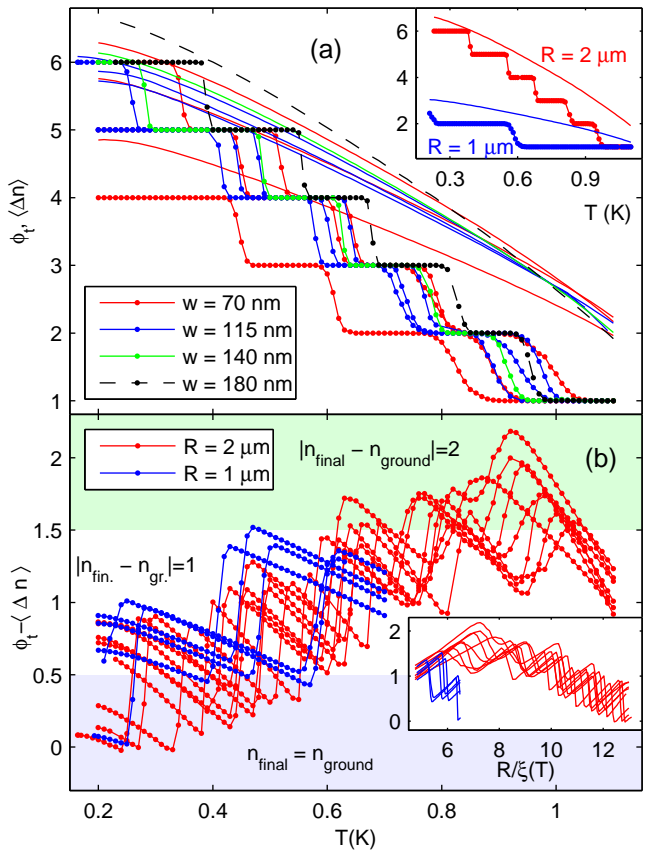}
\caption{\label{fig:all} \label{fig:scale} (Color online) (a) Mean
transition point $\phi_t$ (lines only) and  
$\langle \Delta n\rangle$ (lines with dots, each dot representing one
dataset of 40 field sweeps).  The inset shows the same data for just
the ring from Fig. \ref{fig:prob} and a $R$ = 1 $\mu$m ring.  
(b) $\phi_t$ -  $\langle \Delta n \rangle$ as a function
of temperature for all 13 rings. 
The three shaded regions indicate the difference
$|n_f - n_{ground}|$ between the $n$ of the final and the ground
state. Since $\phi_t - \langle \Delta n \rangle = \pm (\varphi_t - n_f)$, 
it encodes both $|n_f - n_{ground}|$, and the dependence of the
final state on how far from an integer flux bias point a transition
occurs.  Comparison with the inset shows that $\phi_t - \langle \Delta
n \rangle$ depends directly on the temperature rather than $R/\xi(T)$.  
}
\end{figure}

We have extracted the position $\varphi_t$ of each transition and $n$ for each 
branch, and computed $\phi_t(T) \equiv \langle |\varphi_t - n_i| \rangle$ 
by averaging over all 
transitions in each set of field sweeps. $n_i$ and $n_f$ are the $n$ before
and after each transition, as illustrated in Fig.  \ref{fig:sweeps}(d).
The  $\le$ 50 m$\Phi_0$ rms-scatter of $|\varphi_t-n_i|$ around $\phi_t$,  
which is partly due to sensor vibrations, is negligible for our purpose.
The observed frequencies of $\Delta n \equiv |n_i - n_f|$ values yield 
a probability distribution as a function of $T$ for each ring, 
as shown in Fig. \ref{fig:prob}(e). Since we find at most two 
consecutive integer values of $\Delta  n$ at any given $T$, 
the distribution is fully characterized by its 
mean $\langle \Delta n\rangle(T)$ [Fig. \ref{fig:all}].

A GL stability analysis neglecting screening effects and fluctuations predicts 
$\phi_t = \sqrt{R^2 /\xi_{GL}^2 + 1/2}/ \sqrt{3} \approx R/\sqrt{3}\xi(T)$
\cite{VodolazovDY:Mulfja}.
In practice, $\phi_t$ as shown in Fig. \ref{fig:all} varies significantly 
between nominally identical rings, and is up to 30 \% smaller than the 
measured value of $R/\sqrt{3}\xi(T)$, with the largest deviation for 
the smallest $w$.
The variations suggests that imperfections contribute significantly to this 
discrepancy, but a deviation from GL theory far below $T_c$ and thermal 
activation or tunneling may also play a role. 
The temperatures at which $\Delta n$ changes also show significant scatter.
However, they correlate strongly with the value of $\phi_t$
at that point for similar $T$.  
Consequently, the curves of $\phi_t(T) - \langle \Delta n\rangle(T)
= \pm (\varphi_t - n_f)$ 
from all rings [Fig. \ref{fig:scale}] approximately collapse into
a band whose width of $\approx 1$ is largely due to the discreteness of 
$\langle \Delta n\rangle$. 
Below 0.5 $T_c$, $\phi_t - \langle \Delta n\rangle < 1/2$ can
occur, which implies
that the final state is the ground state with $|\varphi_t - n_f| < 1/2$.
The lower $T$, the larger is the range of $\varphi_t$ over which the 
ground state is reached.
The absence of a clear $w$ dependence of 
$\phi_t - \langle \Delta n\rangle$ vs. $T$ indicates that the effect
of the self inductance and the thermal 
activation barrier, both of which are proportional to $w d$, is small.

The results show no significant dependence on sweep 
frequency from 0.6 to 9.6 Hz.
One concern is heating from the $\sim$5 GHz, $\sim$5 m$\Phi_0$
Josephson oscillation applied to the rings by the SQUID.
Varying the pickup loop--ring inductance $M_{coup}$ by a factor 5 by 
retracting the SQUID about 2 $\mu$m had no significant effect. 
However, when the 
sensor chip touched the sample substrate, $\phi_t(T)$ flattened slightly 
 below 0.35 K, changing by 0.1 at 0.2 K.
Since this effect is negligible, 
the data was acquired 
with the SQUID touching the sample to avoid variations of $M_{coup}$ due to
scanner drift and vibrations.

We now turn to discussing our results in terms of the  TDGL equations for 
gapped superconductors:
\begin{subequations}
\label{eq:TDGLgapped}
\begin{eqnarray}
\lefteqn{
\frac u {\sqrt{1 + \gamma^2 |\psi|^2}}\left( \frac{\partial}{\partial t} 
+ i V  + \frac {\gamma^2} 2 \frac {\partial |\psi|^2}{\partial t}\right)\psi}
\hspace{2cm} \nonumber \\
&=& (\nabla - i \mathbf{A})^2\psi + (1-|\psi|^2)\psi \label{eq:TDGLgappedA} \\ 
\nabla^2 V &=& \nabla\cdot\mathrm{Re}(\psi^*(-i\nabla + \mathbf A)\psi),
\end{eqnarray}
\end{subequations}
where $\gamma = 2 \tau_E \Delta_0(T)/\hbar$. 
$\Delta_0(T)$ is the equilibrium gap, $V$ the electrostatic potential, 
and $\tau_E$ the inelastic electron-phonon scattering time.
Without magnetic impurities, theory predicts $u = \pi^4/14 \zeta(3) = 5.79$
\cite{KramerL:Thedcs}. 
Eqs. (\ref{eq:TDGLgapped}) are written in dimensionless 
variables, as defined in Ref. \cite{MichotteS:Conops}. 
We only note that the unit of time is 
$\tau_{GL} = 2 k_B T \hbar/\Delta_0^2$, and 
the dimensionless ring circumference 
 is $2 \pi R/\xi(T)$. 
The latter might lead to an indirect $T$-dependence. However, 
Fig. \ref{fig:scale}(b)
shows only a weak $R$-dependence of the range of possible 
$\phi_t(T) - \langle \Delta n\rangle(T)$, whereas plotting 
$\phi_t - \langle \Delta n\rangle$ against $R/\xi(T)$ 
(inset) gives two distinct bands. Thus, the 
$T$-dependence of  $\gamma$ is more important
than that of $R/\xi(T)$, to the extent that Eqs. (\ref{eq:TDGLgapped})
are applicable. 

The mechanism underlying multiple PS's in mesoscopic rings was 
analyzed in Refs. \cite{VodolazovDY:Dyntbm, VodolazovDY:Mulfja}, 
based on numerical solutions of Eqs. \ref{eq:TDGLgapped} with $\gamma = 0$.
Ref. \cite{VodolazovDY:Dyntbm} identifies two
timescales: the duration of a single PS, $\tau_\phi$, and 
the relaxation time $\tau_{|\psi|}$ of $|\psi|$.
If $\tau_\phi \lesssim \tau_{|\psi|}$, the phase may unwind by several 
$2\pi$ before  $|\psi|$ can recover after the first PS.
Further analysis for wires with finite $\gamma$ \cite{MichotteS:Conops} 
shows that $\tau_\phi$ is related to the voltage drop $\Delta V = I R_{PS}$
across the PS center via the Josephson relation: $\tau_\phi = \Phi_0/\Delta V$,
where $R_{PS}$ is the resistance of the region around the PS
over which the electric field decays and the QP current is 
converted to supercurrent. The  
extent of this region is given by the charge imbalance 
length $\Lambda_{Q^*} \propto  \sqrt{\gamma / u} \xi(T)$. 
One obtains $\tau_\phi \propto  \sqrt{u / \gamma } \tau_{GL}$ and
$\tau_{|\psi|} \propto \gamma u \tau_{GL}$. 
The above expressions are valid for $\gamma \gg 1$. For $\gamma \le 1$, one 
should replace $\gamma$ with 1. 

If treating $u$  as a phenomenological adjustable parameter and setting
$\gamma = 0$, one thus finds that 
$\tau_\phi/\tau_{|\psi|}$ decreases with increasing $u$, so that a larger 
$u$ leads to larger $\Delta n$ \cite{VodolazovDY:Dyntbm}. 
Reasonable agreement with a previous 
experiment at 0.4 K was obtained for $u$ = 48  \cite{VodolazovDY:Mulfja}.
However, $u \gg 6$ is 
inconsistent with microscopic theory and experiments in the 
resistive state of quasi-1D superconducting wires 
\cite{IvlevBI:Elecar, TidecksR:Curnpi}.
Those show that except very close to $T_c$, 
$\xi < \Lambda_{Q^*} \propto |T-T_c|^{-1/4}$. 
Some experiments even found a constant $\Lambda_{Q^*}$. 
When using Eqs. (\ref{eq:TDGLgapped}) with $\gamma = 0$,
this implies $u < 1$ and that $u$ decreases with decreasing $T$ 
\cite{VodolazovDY:Dyntbm}. For example, Eqs. (\ref{eq:TDGLgapped}) 
with $u = 0.01$ and $\gamma =0 $ have been used to model the resistive state
\cite{IvlevBI:Elecar}.

This inconsistency in the effective $u$ arises because
$\gamma$ is negligible only in the gapless limit, 
where phonon (or magnetic impurity) induced
pair breaking causes a fast reaction 
of the QP population to order parameter variations.
For pure Al, this limit only applies for  $T > T_c - 10^{-10}$ K.
An increase of 
$\gamma \propto \sqrt{|T-T_c|}$ with decreasing $T$ on the other hand
both reduces $\tau_\phi/\tau_{|\psi|}$ and
increases $\Lambda_{Q^*}/\xi(T)$, 
in qualitative agreement with both ring and resistive-state experiments.
However, Eqs. (\ref{eq:TDGLgapped}) neglect heating and QP diffusion,
which is valid for $\tau_\phi, \tau_{|\psi|} \ll  \tau_E$.
Thus, even accounting for the slow charge imbalance relaxation
by allowing $\gamma \ne 0$ 
only extends the  theoretical  validity to $T_c -T < 0.1$ mK for Al 
\cite{Watts-TobinRJ:Nontdc} \footnote{For some common 
low-$T_c$ superconductors with shorter $\tau_E$, 
Eqs. (\ref{eq:TDGLgapped}) hold over a much larger temperature range.}. 

The observation that $\varphi_t -n_f$ is nearly independent of $R$
implies that the current after a transition is approximately inversely 
proportional to $R$.
For wires carrying a quasi-dc bias current $I$ on the other hand, phase 
slipping stops once $I$ drops below some limit $I_{c1}$, at which
$\tau_\phi = \Phi_0/R_{PS}I_{c1} \approx \tau_{|\psi|}$ 
\cite{MichotteS:Conops}.
Within the TDGL picture discussed above, this finite length effect
could be explained if the normal-like length determining $R_{PS}$
is set by circumference rather than  $\Lambda_{Q^*}$, which is formally
plausible since  $2 \Lambda_{Q^*} > 2 \pi R$ at $T$ = 0.
However, QP's can only diffuse a distance $\sqrt{D \tau_\phi} \ll \pi R$ 
during a single PS, so that it seems questionable if the
whole ring can contribute to $R_{PS}$.
Alternatively, the local reduction of the critical current at the PS center
could increase as more PS's occur, so that larger rings holding more fluxoids 
have a smaller final current.
Well below $T_c$, where $\Delta_0 \gg k_B T$, the electronic heat capacity 
$c(T)$ is very small, while
the kinetic energy density is large.
Therefore, dissipating the latter in a PS will lead to 
significant heating, i.e. excitation of the energy mode of the
QP population, which increases for larger $R$. Thermalization with the lattice 
occurs on the scale of $\tau_E \ge 50$ ns \cite{Watts-TobinRJ:Nontdc}, whereas
assuming that $R_{PS}$ is at least that of a ring section of length $\xi$
gives $\tau_\phi \lesssim 10$ ps.
Thus, the electrons are approximately  a closed system.
If the kinetic energy was converted to homogeneous heating, the
resulting electronic temperature would be given, very roughly,
by max($T$, 0.8 K), because $c(T)$ is small for $T \lessapprox 0.5 T_c$.
At this temperature, $\phi_t$ is still rather large, so that uniform 
heating would lead to smaller $\Delta n$ than observed. However,
given the short diffusion length $\sqrt{D \tau_\phi} \sim \xi \ll 2 \pi R$,
hot QP's will remain localized so that diffusive cooling may affect 
$\tau_{|\psi|}$.

A general description of time dependent superconductivity, 
similar to the Usadel equations for the static case, was derived 
in Ref. \cite{Watts-TobinRJ:Nontdc}. 
Unfortunately, the resulting system of equations for 
$\Delta$, the Green's functions and electron 
distribution is rather complex. Approximate solutions for specific cases were
studied in Ref. \cite{BaratoffA:Selndp, SchmidA:Dynpsw}.
Similar computations might allow a direct comparison between our 
data and microscopic theory.
A much simpler first step could be a similar study as in 
\cite{VodolazovDY:Dyntbm} using Eqs. (\ref{eq:TDGLgapped}), but with $\gamma$ 
rather than $u$ as free parameter. 
Heating and QP diffusion could be accounted for through a local effective 
temperature governed by the heat equation.

In conclusion, we have shown experimentally that the final state after
a field induced fluxoid transition in mesoscopic rings approaches the
ground state for $T \rightarrow 0$. 
While localized heating likely is important, our results cannot be explained
with uniform heating.  Time dependent Ginzburg-Landau theory for
gapped superconductors might give an adequate phenomenological model,
however it is not microscopically justified in the experimental
temperature range.
Although our experiment is conceptually rather simple,
a quantitative explanation is intriguingly complex and
probably requires the incorporation of QP diffusion.  Such
a model may also provide insight in localized nonequilibrium effects
on the superconducting dynamics in other cases, such as vortex
motion or flux avalanches.

\acknowledgments{This work was supported primarily by NSF Grant No.
DMR-0507931, with additional support by 
DMR-0216470, ECS-0210877 and PHY-0425897 and by the 
Packard Foundation. Samples were fabricated at the Stanford 
Nanofabrication Facility of NNIN supported by NSF Grant No. ECS-9731293. 
We would like to thank Denis Vodolazov and Jorge Berger for useful discussions 
and comments.
}

\bibliography{two_comp_bibdata,pc_bibdata,multijump_bibdata}

\end{document}